\documentclass[a4paper, preprint, jcp]{revtex4}
\usepackage{amsmath}
\usepackage{amsfonts}
\usepackage{graphicx}
\usepackage{color}
\usepackage{lscape}
\usepackage{amssymb}

\begin{document}

\title{Dynamical simulations of charged soliton transport in conjugated polymers with the inclusion of electron-electron interactions}
\author{Haibo Ma}
\email{haiboma@physik.rwth-aachen.de}
\author{Ulrich Schollw\"{o}ck}
\email{scholl@physik.rwth-aachen.de} \affiliation{Institut f\"{u}r
Theoretische Physik C, RWTH Aachen University, D-52056 Aachen,
Germany}
\date{Latest revised on \today}

\begin{abstract}

We present numerical studies of the transport dynamics of a charged
soliton in conjugated polymers under the influence of an external
time-dependent electric field. All relevant electron-phonon and
electron-electron interactions are nearly fully taken into account
by simulating the monomer displacements with classical molecular
dynamics (MD) and evolving the wavefunction for the $\pi$ electrons
by virtue of the adaptive time-dependent density matrix
renormalization group (TDDMRG) simultaneously and nonadiabatically. It is found that after a smooth
turn-on of the external electric field the charged soliton is
accelerated at first up to a stationary constant velocity as one
entity consisting of both the charge and the lattice deformation. An
ohmic region (6 mV/$\text{\AA}$ $\leq E_0\leq$ 12 mV/$\text{\AA}$)
where the stationary velocity increases linearly with the electric
field strength is observed. The relationship between
electron-electron interactions and charged soliton transport is also
investigated in detail. We find that the dependence of the
stationary velocity of a charged soliton on the on-site Coulomb
interactions $U$ and the nearest-neighbor interactions $V$ is due to
the extent of delocalization of the charged soliton defect.

\end{abstract}

\maketitle
\section{Introduction}
For more than three decades, there has been a great number of both
academic and industrial interest in \textit{trans}-polyacetylene
(PA), since the discovery that its electrical conductivity can be
improved greatly through charge doping or photo-excitation
\cite{Chiang77, Shirakawa77, Chiang78}. It is now widely accepted
that the enhanced conductivity of \textit{trans}-PA is due to
its inherent tendency to show solitonic behavior. Because there
exists a twofold degeneracy of ground state energy in \textit{trans}-PA
distinguished by the positions of the double and single bonds,
the soliton excitations take the form of domain walls separating
different districts of opposite single and double bonds alternation
patterns. The reduction or the oxidation of the \textit{trans}-PA chain will result in polaron defect and/or spinless charged soliton defect. At least at low doping levels, conductivity of \textit{trans}-PA is thought to result from the motion of these quasiparticles.\cite{Su79, Su80, Heeger88, Heeger01, Heeger01_2, Baeriswyl92, Barford05}

Obviously, it is of great importance to understand the underlying
mechanisms of charge carrier transport under an external electric
field in order to modulate or devise new functional materials based
on \textit{trans}-PA. There have been extensive theoretical studies
of the dynamics of solitons in \textit{trans}-PA under the influence
of external forces \cite{Su79, Su80, Bishop84, Heeger88,
Johansson02} based on the Su-Schrieffer-Heeger (SSH) model
\cite{Su79, Su80}, an improved H\"{u}ckel molecular orbital model,
in which $\pi$ electrons are coupled to distortions in the polymer
backbone by the electron-phonon interaction. In order to improve the
situation that there are no electron-electron  interactions included
in SSH calculations, F\"{o}rner \textit{et al.} \cite{Forner88, Forner98}
performed Pariser-Parr-Pople (PPP) simulations for the soliton
dynamics in \textit{trans}-PA at the self-consistent field (SCF)
level and found that the soliton width decreased with the inclusion
of electron-electron interactions and that the soliton velocity
($v_s$) is smaller than that found in SSH calculations. By using an \textit{ab initio} molecular dynamic with a single spin-unrestricted determinant, Champagne \textit{et al.} characterized the motion of carbon and hydrogen atoms in a small polyacetylene chain with a positively charged soliton defect driven by an external electric field and found that the motion consists of dominant longitudinal CH displacements combined with less prominent H wags and CH stretches.\cite{Champagne97}

However, recently a lot of theoretical calculations of the static properties of
conjugated polymers have shown that the electron correlation effect
plays a very important role in determining the behavior of the
charge carriers in conjugated polymers.\cite{Yonemitsu88, Sim91,
Suhai92, Villar92, Rodriguez-Monge95, Hirata95, Bally92, Fulscher95,
Guo97, Perpete99, Fonseca01, Oliveira03, Champagne04, Monev05, Baeriswyl92, Barford05} For example, it is found that the soliton width depends strongly upon the
inclusion of the electron correlation effects and the electron
correlation effects are substantial for characterizing the geometrical and charge features of $\pi$-conjugated chains bearing a
charged soliton defect without counterion.\cite{Champagne04} Therefore, theoretical simulations for
soliton dynamics with electron correlations considered, which step
beyond the SCF level, are highly desirable. However, the calculations for
large polymer systems with traditional advanced electron-correlation
methods such as configuration interaction method (CI), multi-configuration self-consistent field method (MCSCF), many-body Moller-Plesset perturbation theory (MPn) and coupled cluster method (CC) are currently still not feasible due to the huge computational
costs. Fortunately, the density-matrix renormalization group (DMRG)
method, firstly proposed by White in 1992 \cite{White92, White93,
Schollwock05}, can be used instead: it has been demonstrated and
widely accepted that at economic computational cost the accuracy of DMRG is very close to exact full configuration interaction (FCI)
calculations for one-dimensional (1D) or quasi-1D systems, where DMRG is the most performing numerical method currently available. Since Pang and Liang introduced the DMRG method into the studies of conjugated polymers with an alternating Hubbard model in 1995 \cite{Pang95}, a lot of interesting papers by many groups focusing on the static investigations of conjugated polymers with the application of DMRG within various empirical or semi-empirical models have been published \cite{Wen96, Lepetit97, Yaron98, Boman98, Kuwabara98, Shuai98, Shuai98_2, Bursill99, Zhang00, Barford01, Barford01_2, Race01, Barford02, Barford02_2, Raghu02, Race03, Ma04, Moore05, Ma05, Ma05_2, Yan05, Ma06, Hu07, Ma07}.

In this paper, combining the SSH and the extended Hubbard model
(EHM), we simulate the motion of the charged soliton in
\textit{trans}-PA under an applied external electric field
based on a nonadiabatic approach, in which we evolve the
$\pi$-electron wavefunction using a time-evolution scheme for DMRG
(the adaptive time-dependent density matrix renormalization group
(TDDMRG) \cite{White04, Daley04}) and simulate the motion of the
nuclei in the lattice backbone by classical molecular dynamics (MD)
under the couplings with the $\pi$-electron part simultaneously,
denoted for convenience as TDDMRG/MD. The aim of this paper is to give an exhaustive
picture of charged soliton transport in \textit{trans}-PA at a
theoretical level with all relevant electron-electron interactions
and correlations included and to show how the on-site Coulomb
interactions and nearest-neighbor electron-electron interactions
influence the behavior of charged soliton transport in
\textit{trans}-PA.
\section{Model and Methodology}

\subsection{Model}
We use the well-known and widely used SSH Hamiltonian \cite{Su79,
Su80} combined with the extended Hubbard model (EHM), and include
the external electric field by an additional term:
\begin{equation}\label{H}
    H(t)=H_{el}+H_{E(t)}+H_{latt}
\end{equation}
This Hamiltonian is time-dependent, because the electric field $E(t)$ is controlled explicitly by time.

The $\pi$-electron part includes both the electron-phonon and the
electron-electron interactions,
\begin{equation}\label{H_el}
\begin{split}
    H_{el}=&-\sum_{n,\sigma}t_{n,n+1}(c_{n+1,\sigma}^{+}c_{n,\sigma}+h.c.)\\
& +\frac{U}{2}\sum_{n,\sigma}(c_{n,\sigma}^{+}c_{n,\sigma}-\frac{1}{2})(c_{n,-\sigma}^{+}c_{n,-\sigma}-\frac{1}{2})\\
&
+V\sum_{n,\sigma,\sigma'}(c_{n,\sigma}^{+}c_{n,\sigma}-\frac{1}{2})(c_{n+1,\sigma'}^{+}c_{n+1,\sigma'}-\frac{1}{2})
\end{split}
\end{equation}
where $t_{n,n+1}$ is the hopping integral between the $n$-th site
and the ($n+1$)-th site, while $U$ is the on-site Coulomb interaction
and $V$ denotes the nearest-neighbor electron-electron interaction.
Because the distortions of the lattice backbone are always within a
certain limited extent, one can adopt a linear relationship between
the hopping integral and the lattice displacements as
$t_{n,n+1}=t_0-\alpha(u_{n+1}-u_n)$ \cite{Su79, Su80}, where $t_0$
is the hopping integral for zero displacement, $u_n$ the lattice
displacement of the $n$th CH monomer, and $\alpha$ is the electron-phonon
coupling.

Because the atoms move much slower than the electrons, we treat lattice backbone classically with the Hamiltonian
\begin{equation}\label{H_latt}
    H_{latt}=\frac{K}{2}\sum_{n}(u_{n+1}-u_n)^2+\frac{M}{2}\sum_{n}\dot{u}_n^2 \text{    },
\end{equation}
where $K$ is the elastic constant and $M$ is the mass of a CH monomer.

The electric field $E(t)$ directed along the backbone chain is uniform over
the entire system. The field which is constant after a smooth
turn-on is chosen to be
  \begin{equation}\label{eq:t2}
    E(t)=\begin{cases}
    E_0exp[-(t-T_C)^2/T_W^2],    &\text{for $t< T_C$, } \\
    E_0,    &\text{for $t\geq T_C$, }
    \end{cases}
  \end{equation}
where $T_C$, $T_W$  and $E_0$ are the center, width and strength of
the half Gaussian pulse. This field gives the following contribution to the
Hamiltonian:
\begin{equation}\label{H_Et}
    H_{E(t)}=\vert e\vert\sum_{n,\sigma}(na+u_n)(c_{n,\sigma}^{+}c_{n,\sigma}-\frac{1}{2})E(t)
\end{equation}
where $e$ is the electron charge and $a$ is the lattice constant.
The model parameters are those generally chosen according to Su \textit{et al}'s pioneer work \cite{Su80}:
$t_0$=2.5 eV, $\alpha$=4.1 eV/\AA, $K$=21 eV/\AA$^2$, $M$=1349.14
eVfs$^2$/\AA$^2$, $a$=1.22 \AA. We notice that there are some debates on the accuracy of some of the parameters. For example, Ehrenfreund, \textit{et al.} proposed to use $K$=46 eV/\AA$^2$ instead of 21 eV/\AA$^2$ to fit the resonant Raman scattering experiment \cite{Ehrenfreund87}, and an additional linear term was suggested to be included into to Eq.~\ref{H_latt} for the more proper description of the elastic energy of the C-C $\sigma$ backbone \cite{Baeriswyl92} since $a$ is assumed to be the equilibrium lattice spacing of the undimized chain, including both $\sigma$- and $\pi$-bonding. Fur the purpose of making contact to other theoretical studies on this problem 
\cite{Rakhmanova99, Rakhmanova00, Johansson01, Johansson02, Basko02, Johansson04, Fu06, Zhao08, Ma08}, we choose the current parameter schemes. In order to prevent the contraction of the chain, which might result from our simplified lattice elastic energy form, we supplement a constraint of fixed total length of the chain. The
values of $T_C$ and $T_W$ ($T_C$=30 fs and $T_W$=25 fs) are taken
from the paper of Fu \textit{et al}.\cite{Fu06}

\subsection{the TDDMRG/MD method}
Before moving to the dynamical simulation, we
determine the static electronic structure of the system in the
absence of the external electric field by virtue of DMRG. The DMRG
method \cite{White92, White93} has emerged over the last decade as
the most powerful method for the simulation of strongly correlated
1D quantum systems. We would like to
refer readers to recent reviews \cite{Schollwock05, Schollwock07, Schollwock07_2} for the technical details of DMRG. During the DMRG
calculations, we utilize energy minimization methods \cite{Chadi78,
Payne92} to optimize the initial lattice configuration \{$u_n$\} for
the chain backbone with the energy gradients being calculated according
to the Hellmann-Feynman theorem \cite{Hellmann37, Feynman39}. 

\begin{figure}
\begin{center}
\includegraphics[width =10 cm ]{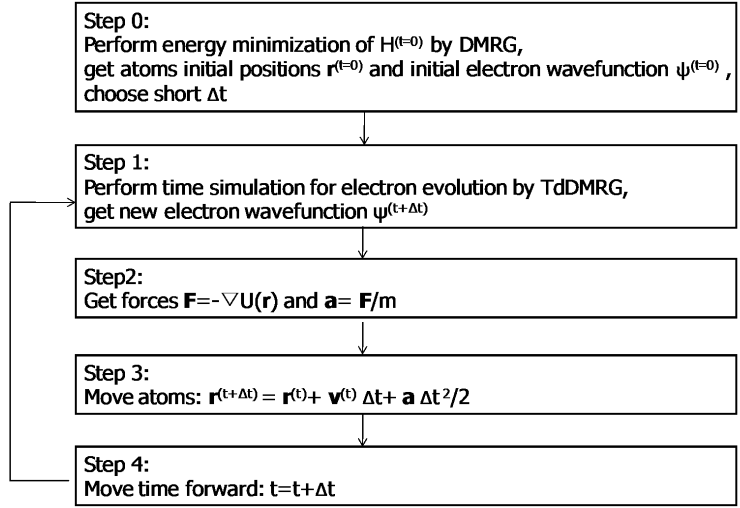}
\caption{\label{fig:MD} Procedures for the TDDMRG/MD method.}
\end{center}
\end{figure}

After the lattice configuration and the
$\pi$-electron wavefunction for the initial state are set, we turn
on the external electric field smoothly and started the real-time
simulation of the dynamic process of the charged soliton transport
along the model chain by the nonadiabatic TDDMRG/MD method. The main
idea of the TDDMRG/MD method is to iteratively repeat the procedures of evolving the $\pi$-electron part by
the adaptive TDDMRG and moving the backbone part by classical MD
simultaneously and nonadiabatically. The flowchart of the procedures for the TDDMRG/MD
method is illustrated in Fig.~\ref{fig:MD}.

The adaptive TDDMRG
method \cite{White04, Daley04}, which is an
extension of standard DMRG using Vidal's time-evolving block-decimation (TEBD) algorithm
\cite{Vidal04}, is a recently developed very efficient numerical method to solve the time-dependent Schr\"{o}dinger equation
\begin{equation}\label{Sch}
\begin{split}
    &H(t)\Psi(t)=i\hbar\frac{\partial \Psi(t)}{\partial t}\\
&\Psi(t+\Delta t)=U(\Delta t)\Psi(t)\text{   with }U(\Delta
t)=\text{exp}(-iH(t)\Delta t/\hbar)\text{    }.
\end{split}
\end{equation}
Time evolution in the adaptive TDDMRG is generated using the
Trotter-Suzuki decomposition of the time-evolution operator
$U(\Delta t)$ of Eq.~(\ref{Sch}). Since the Hamiltonian operator for
the $\pi$-electron part of Eq.~(\ref{H_el}) can be decomposed into a
sum of local terms $h_n$ that live only on neighboring sites $n$ and
$n+1$, $U(\Delta t)$ can be approximated by an $n$th-order
Trotter-Suzuki decomposition \cite{Suzuki76}, e.g., to second order,
\begin{equation}\label{Un}
    U(\Delta t)=\prod_{n\in even}U_n(\Delta t/2)\prod_{n\in odd}U_n(\Delta t)\prod_{n\in even}U_n(\Delta t/2)+O(\Delta t^3)
\end{equation}
The $U_n(\Delta t)$ are the infinitesimal time-evolution operators
exp$(-ih_n(t)\Delta t/\hbar)$ on the bond linking sites $n$ and
$n+1$. The ordering within the even or odd products doesn't matter,
because ``even'' and ``odd'' operators commute among themselves.
Then we can incorporate the applications of these operators
$U_n(\Delta t)$ successively to some state $|\psi\rangle$ into
finite-system DMRG sweeps \cite{White92, White93, Schollwock05}.
Each operator $U_n(\Delta t)$ is applied at a finite-system DMRG
step with sites $n$ and $n+1$ being the active sites, i.e., where
sites $n$ and $n+1$ are represented without truncation. For example,
in a $L$-site 1D chain, at a finite-system DMRG step with $n$ and
$n+1$ being the active sites, we can use Eq.~(\ref{MPS}) to describe
a DMRG state in the form of matrix product states (MPS)
\cite{Klumper91, Fannes92, Klumper92, Derrida93, Ian07} (also known
as finitely correlated states),
\begin{equation}\label{MPS}
    |\psi\rangle=\sum_{\sigma_1...\sigma_L}A^1[\sigma_1]A^2[\sigma_2]...A^L[\sigma_L]|\sigma_1...\sigma_L\rangle,
\end{equation}
where $\{|\sigma_i\rangle\}$ denotes the state space for site $i$, matrices $A^i$ are for site $i$, and the first and last $A$-matrices are taken to be row and column vectors. Then it is easy to
describe the new state after the operation of $U_n(\Delta t)$
\begin{equation}\label{Psin}
    U_n(\Delta t)\vert\psi\rangle=\sum_{\sigma_n'\sigma_{n+1}'}\sum_{\sigma_1...\sigma_L}U_n(\Delta t)_{\sigma_n'\sigma_{n+1}',\sigma_n\sigma_{n+1}}A^1[\sigma_1]...A^n[\sigma_n']A^{n+1}[\sigma_{n+1}']...A^L[\sigma_L]|\sigma_1...\sigma_n'\sigma_{n+1}'...\sigma_L\rangle
\end{equation}
without any additional error, because $U_n(\Delta t)$ only acts on
the part of Hilbert space
($\vert\sigma_n\rangle\vert\sigma_{n+1}\rangle$) which is exactly
represented. Similar to conventional DMRG, in order to prevent the
exponential growth of matrix dimension for continuing the
finite-system sweep, DMRG truncations must be carried out. But
differently, the adaptive TDDMRG uses $U_n(\Delta
t)\vert\Psi\rangle$ as the target state instead of
$\vert\Psi\rangle$ to build the reduced density matrix. At the end
of several sweeps, when all the local time-evolution operators have
been applied successively to $\Psi(t)$, one can get the
$\pi$-electron wavefunction $\Psi(t+\Delta t)$ for the new time
$t+\Delta t$, and continue to longer times. In the context of 1D
electronic and bosonic systems, the adaptive TDDMRG has been found
to be a highly reliable method, for example in the context of
magnetization dynamics \cite{Gobert05} of spin-charge separation
\cite{Kollath05, Kleine08} or far-from equilibrium dynamics of
ultracold  bosonic atoms \cite{Cramer08}.

During the TDDMRG/MD simulations, after we evolve
the $\pi$-electron wavefunction from $t$ to $t+\Delta t$ and get
$\Psi(t+\Delta t)$ using the adaptive TDDMRG technique, we start
classical MD simulations for the motion of atom nuclei in the chain
backbone. The force acting on the $n$th site is equal to the
negative energy gradient of $n$th site's displacement. Here, we take
the non-Born-Oppenheimer effect into account and calculate these
forces in a nonadiabatic way, as in Eq.~(\ref{F1}).
\begin{equation}\label{F1}
    M\ddot{u}_n(t)=-\frac{\partial E}{\partial u_n}=-\langle\Psi(t)\vert\frac{\partial H(t)}{\partial u_n}\vert\Psi(t)\rangle
\end{equation}
In the case of our SSH-EHM Hamiltonian, we can calculate these forces as
\begin{equation}\label{F2}
\begin{split}
    M\ddot{u}_n(t)=&-K[2u_n-u_{n+1}-u_{n-1}]\\
&+\vert e\vert E(t)[1-\langle\Psi(t)\vert\sum_{\sigma}c_{n,\sigma}^{+}c_{n,\sigma}\vert\Psi(t)\rangle]\\
&+\alpha\sum_{\sigma}[\langle\Psi(t)\vert
c_{n+1,\sigma}^{+}c_{n,\sigma}+h.c.\vert\Psi(t)\rangle-\langle\Psi(t)\vert
c_{n,\sigma}^{+}c_{n-1,\sigma}+h.c.\vert\Psi(t)\rangle].
\end{split}
\end{equation}
Once we obtain the forces acting on the atom nuclei, we can
simulate the movement of CH monomers according to Newton's laws of motion.

When the new lattice configuration \{$u_n$\} for
the backbone chain is determined, we move the time forward and go
back to the TDDMRG step to evolve the $\pi$-electron wavefunction
and then repeat this process as long as we need or want. Upon
repeating this process again and again, we can simulate a long time
vibration process of the chain backbone, while the $\pi$-electron
part is also well simulated quatum mechanically in real time by
virtue of the adaptive TDDMRG.

\section{Results and discussion}
To investigate the dynamic properties of solitons, we simulate the
charged soliton transport process in a single model conjugated
polymer chain under a uniform external electric field. In order to
present the twofold degeneracy of the ground state energy in soliton
defect, the finite-sized model chain is specified to have an odd
number of sites. In our calculations, we simulate a
positively charged model chain containing $N=101$
CH monomers and $N_e=100$ $\pi$ electrons. Due to
the inherent electron-hole symmetry, our employed parameterized
SSH-EHM model can not be used to distinguish positively and
negatively charged defects. In order to simulate a longer dynamic
process of charged soliton transport within the chain with a certain
length, the charged soliton soliton is initially located in the left
side of the chain instead of being centered in the middle of the
chain as usual. The charged soliton is initially centered around
site 21 through imposing a constraint of reflection symmetry around
site 21. After then, a 160-femtoseconds (fs)
dynamical process, which is generally long enough for the charged
soliton to be transported from the left end to the right end in our
model chain, is simulated by the nonadiabatic TDDMRG/MD method.

Because this is the first time that we perform the dynamical
simulations using MD combined with TDDMRG, we will firstly give a
demonstration that one can achieve reliable results for 1D
correlated quantum systems with appropriate values chosen for two
key parameters. After the submission of this work,
we notice that very recently Zhao \textit{et al} have applied such a
TDDMRG/MD method to the studies of polaronic transport within a
combined model of SSH and Hubbard model and the excellent agreement
of their TDDMRG/MD results with the exact numerical results for a
noninteracting chain ($U$=0, $V$=0) verified the reliability of the
TDDMRG/MD method.\cite{Zhao08} Therefore, here we just mainly focus
on showing the convergence behavior of the numerical results with
different values for the two key parameters and showing how to
choose suitable parameter values for the purpose of achieving
reliable results. The two most important parameters in the
nonadiabatic TDDMRG/MD simulations are the DMRG truncation weight
$\epsilon_{\rho}$ and the time step $\Delta t$.

\begin{figure}
\includegraphics[width =14 cm ]{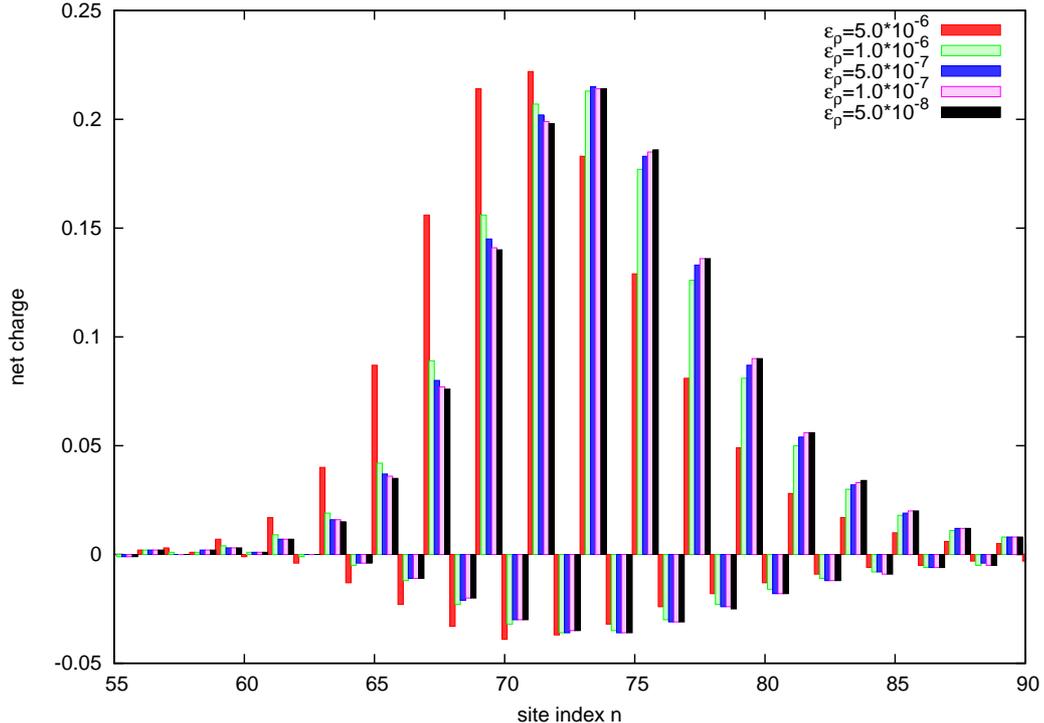}
\caption{\label{fig:Pm} The net charge of a charged soliton as a function of the site
index at $t=160$ fs calculated with different $\epsilon_{\rho}$
values. ($\Delta t$=0.05 fs, $E_0$=6.0 mV/\AA, U=2.0 eV, V=1.0 eV)}
\end{figure}
Fig.~\ref{fig:Pm} shows the net charge distribution pictures at $t=160$
fs calculated by different DMRG truncation weight $\epsilon_{\rho}$ values. As we have
mentioned in the above section, $\epsilon_{\rho}$ is a key parameter for estimating the DMRG truncation
error of the wavefunction and local quantities. So,
$\epsilon_{\rho}$ should be also very vital in our dynamical
simulations by MD combined with the adaptive TDDMRG. As can clearly
seen from Fig.~\ref{fig:Pm}, a big $\epsilon_{\rho}$ value will lead to large deviations from the results obtained by small $\epsilon_{\rho}$ values, and even the peak of the net charge density may be shifted to a totally wrong position. Meanwhile, when we decrease $\epsilon_{\rho}$
value, the net charge values converge gradually; when
$\epsilon_{\rho}$ has been small enough
($\epsilon_{\rho}\leq1.0\times10^{-6}$), the net charge values will
not improve much if we decrease $\epsilon_{\rho}$ value further.
This situation of how $\epsilon_{\rho}$ values influence the final simulated results is very similar to that in standard static DMRG
calculations and it implies that our dynamical simulations may yield reliable results
close to the real case if we control $\epsilon_{\rho}$ to be small enough
($\epsilon_{\rho}\leq1.0\times10^{-6}$).

\begin{figure}
\includegraphics[width =14 cm ]{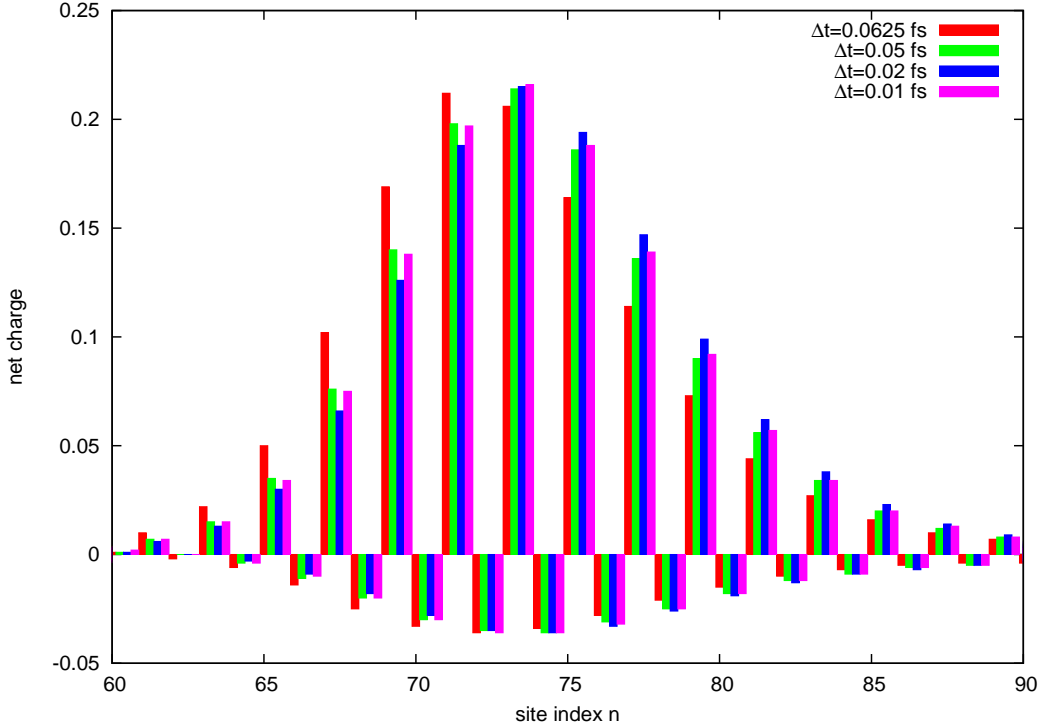}
\caption{\label{fig:Dt} The net charge of a charged soliton as a function of the site
index at $t=160$ fs calculated with different $\Delta t$ values. ($\epsilon_{\rho}=5.0\times10^{-8}$, $E_0$=6.0 mV/\AA, U=2.0 eV, V=1.0 eV)}
\end{figure}
We show the net
charge distribution pictures at $t=160$ fs calculated by different
$\Delta t$ values in Fig.~\ref{fig:Dt}. It is clearly shown
that because of large Trotter error the too large $\Delta t$ value (0.0625 fs) results in apparently different net charge
distribution pictures comparing to the other $\Delta t$ values, while the $\Delta t$
values of 0.05 fs, 0.02 fs and 0.01 fs produce similar charge distribution patterns which are at least qualitatively in good agreement with each other although there are still some slight differences.
Different from the traditional time evolution methods, a smaller time
step $\Delta t$ doesn't always lead to higher accuracy in the
adaptive TDDMRG simulations. Previous studies have elucidated that there are two main error sources in the adaptive TDDMRG:
the Trotter error and the accumulated DMRG truncation error.\cite{Gobert05, Feiguin05}
Although a smaller time step $\Delta t$ will of course result in a much
smaller Trotter error, it requires more DMRG iterations and accordingly it will lead to a
larger accumulated DMRG truncation error. So, one should
choose the time step $\Delta t$ values carefully with a
well-balanced account of these two sorts of errors.
Therefore, we consider that a time step $\Delta t$ value around 0.01$\sim$0.05 fs is
reasonable for our current cases.

\begin{figure}
\includegraphics[width =14 cm]{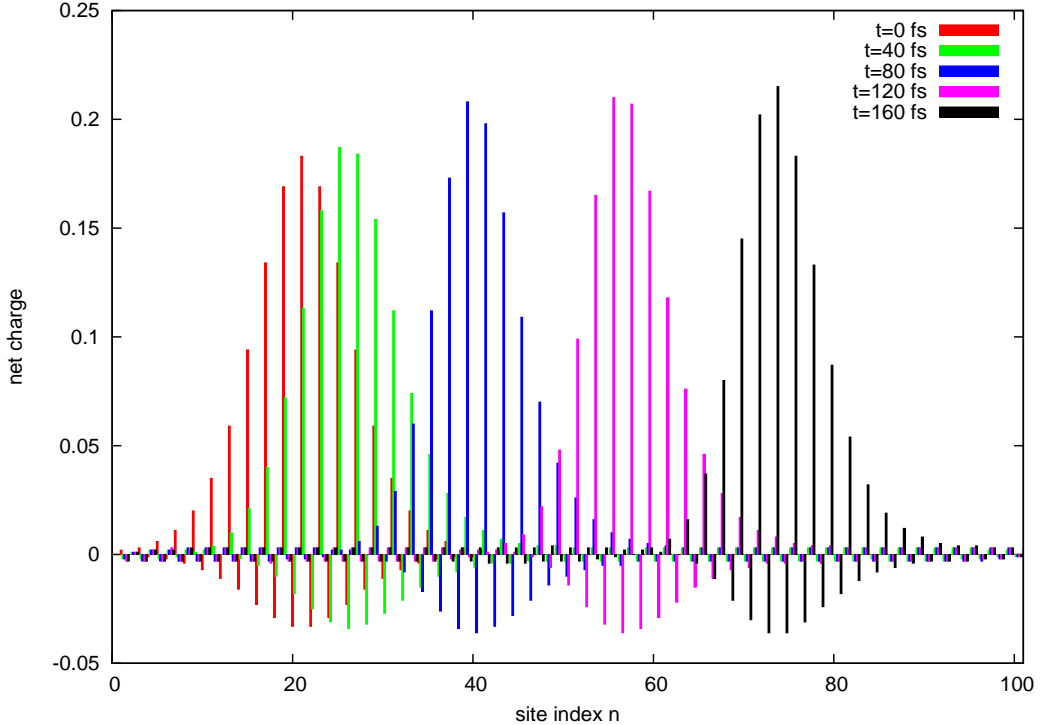}
\caption{\label{fig:c} Time evolution of net charge distribution in a \textit{trans}-PA chain with the transport of a charged soliton under an external electric field. (The charged soliton moves from the left to the right.) ($\Delta t$=0.05 fs, $\epsilon_{\rho}=5.0\times10^{-7}$, $E_0$=6.0 mV/\AA, U=2.0 eV, V=1.0 eV)}
\end{figure}
\begin{figure}
\includegraphics[width =14 cm]{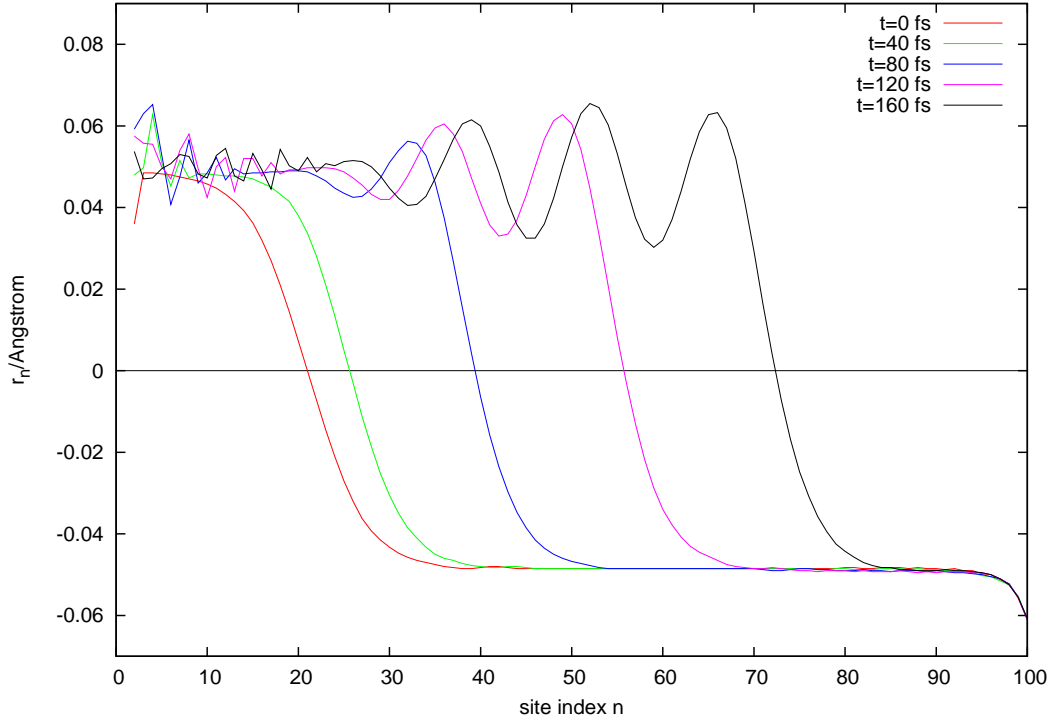}
\caption{\label{fig:g} Time evolution of the staggered bond order parameter $r_n$ in a \textit{trans}-PA chain with the transport of a charged soliton under an external electric field. (The charged soliton moves from the left to the right.) ($\Delta t$=0.05 fs, $\epsilon_{\rho}=5.0\times10^{-7}$, $E_0$=6.0 mV/\AA, U=2.0 eV, V=1.0 eV)}
\end{figure}
We now move on to give the general time evolution
picture of charged soliton transport in \textit{trans}-PA. The
time evolution of net charge distribution and the staggered bond order
parameter $r_n=(-1)^n(2u_n-u_{n+1}-u_{n-1})/4$ in the singly doped
\textit{trans}-PA chain are shown in Fig.~\ref{fig:c} and
Fig.~\ref{fig:g}. It can be clearly seen that, both the net charge and the geometrical distortions don't localize at one CH monomer. On the contrary, the defect is dispersed to several monomers around the soliton center. Actually, defect delocalization is an inherent feature of a soliton and very important for the solitonic transport. One can also clearly see that during the entire time
evolution process the geometrical distortion curve and net charge
distribution shape for the charged soliton defect always stay coupled
and show no dispersion. This implies that the soliton defect is an
inherent feature and the fundamental charge carrier in
\textit{trans}-PA. In Fig.~\ref{fig:g}, we can also see that a long lasting
oscillatory ``tail'' appears behind the soliton defect center. This ``tail'' is generated by the inertia of those monomers to fulfill energy and momentum conservation; they absorb the additional energy, preventing the further increase of the polaron velocity after a stationary value is reached.
\cite{Rakhmanova99, Rakhmanova00, Johansson01, Johansson04}

\begin{figure}
\includegraphics[width =8 cm]{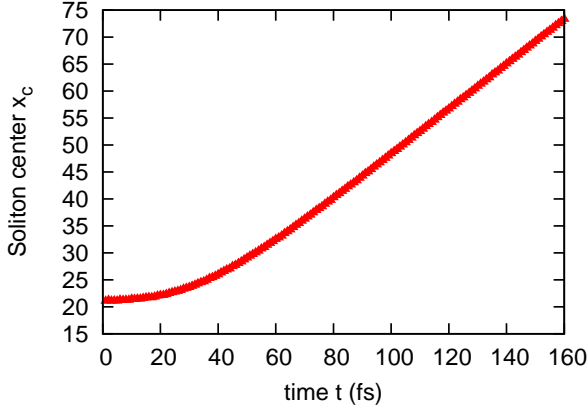}
\caption{\label{fig:Po_t} Temporal evolution of the center position of a charged soliton under an external electric field. ($\Delta t$=0.05 fs, $\epsilon_{\rho}=5.0\times10^{-7}$, $E_0$=6.0 mV/\AA, U=2.0 eV, V=1.0 eV)}
\end{figure}
\begin{figure}
\includegraphics[width =8 cm]{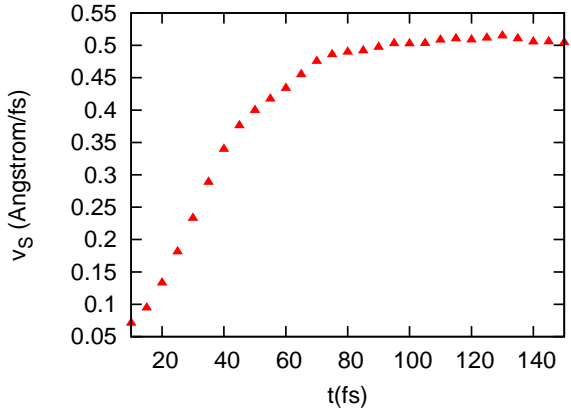}
\caption{\label{fig:v_t} Temporal evolution of the velocity of a charged soliton under an external electric field. ($\Delta t$=0.05 fs, $\epsilon_{\rho}=5.0\times10^{-7}$, $E_0$=6.0 mV/\AA, U=2.0 eV, V=1.0 eV)}
\end{figure}
From Fig.~\ref{fig:c} and Fig.~\ref{fig:g}, one can also find that,
after the external electric field is turned on, the charged soliton is accelerated at
first and then moves with a constant velocity as one entity
consisting of both the charge and the lattice deformation.
Fig.~\ref{fig:Po_t} of the temporal evolution of soliton center position and Fig.~\ref{fig:v_t} of the temporal evolution of soliton velocity
clearly support this observation. Therefore, only the stationary
velocity will be considered for charged soliton transport in the following discussions.

\begin{figure}
\includegraphics[width =8 cm]{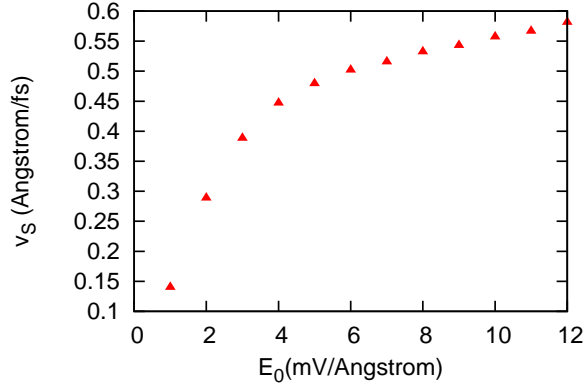}
\caption{\label{fig:E} The stationary velocity of a charged soliton as a function of the
electric field strength. ($\Delta t$=0.05 fs, $\epsilon_{\rho}=5.0\times10^{-7}$, $U$=2.0 eV, $V$=1.0 eV) }
\end{figure}
The dependence of the stationary velocity of a charged soliton $v_s$ on the electric field
strength is shown in Fig.~\ref{fig:E}. We find that $v_s$ increases with increasing electric field strength. Moreover, a very interesting behavior of $v_s$ as a function of the
electric field strength is observed. An ohmic region where $v_s$ increases linearly with
the field strength is found. This ohmic region extends approximately from 6 mV/$\text{\AA}$ to 12 mV/$\text{\AA}$ for the case of $U$=2.0 eV and $V$=1.0 eV. Below 6 mV/$\text{\AA}$ the soliton velocity increases nonlinearly. The saturation of $v_s$ with increasing the
electric field strength is not found.

In order to study the influence of electron-electron interactions
on soliton transport, we focus on the stationary velocity of a charged soliton $v_s$
calculated with different electron-electron interactions under the constraint that
the other parameters are fixed. As should be noticed that, the real conjugated polymer is with weak couplings. Strong couplings will lead to too strong charge polarizations which are unrealistic. Considering that we are only focusing on the
study of real conjugated polymer system, in this work we adopt only the
weak-coupling parameters ($U<2t, V<t$).

\begin{figure}
\includegraphics[width =8cm]{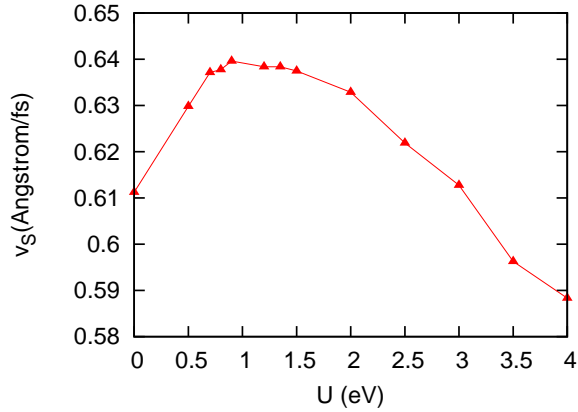}
\caption{\label{fig:v_U} The stationary velocity of a charged soliton as a function of different $U$ values. ($\Delta t$=0.05 fs, $\epsilon_{\rho}=5.0\times10^{-7}$, $E_0$=6.0 mV/\AA, V=0.0 eV)}
\end{figure}
\begin{figure}
\includegraphics[width =14 cm]{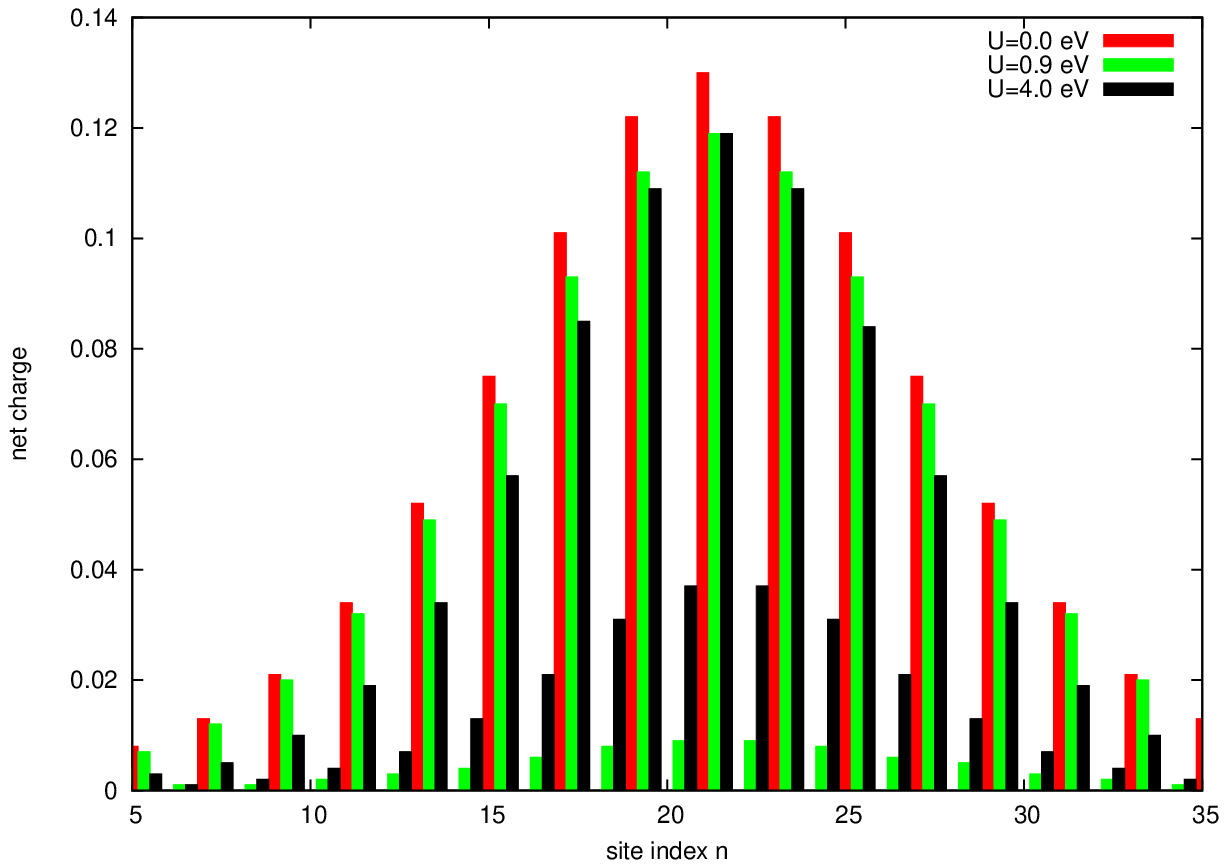}
\caption{\label{fig:c_U} The net charge of a static charged soliton as a function of the site
index calculated with different $U$ values. ($\epsilon_{\rho}=5.0\times10^{-7}$, V=0.0 eV)}
\end{figure}
\begin{figure}
\includegraphics[width =14 cm]{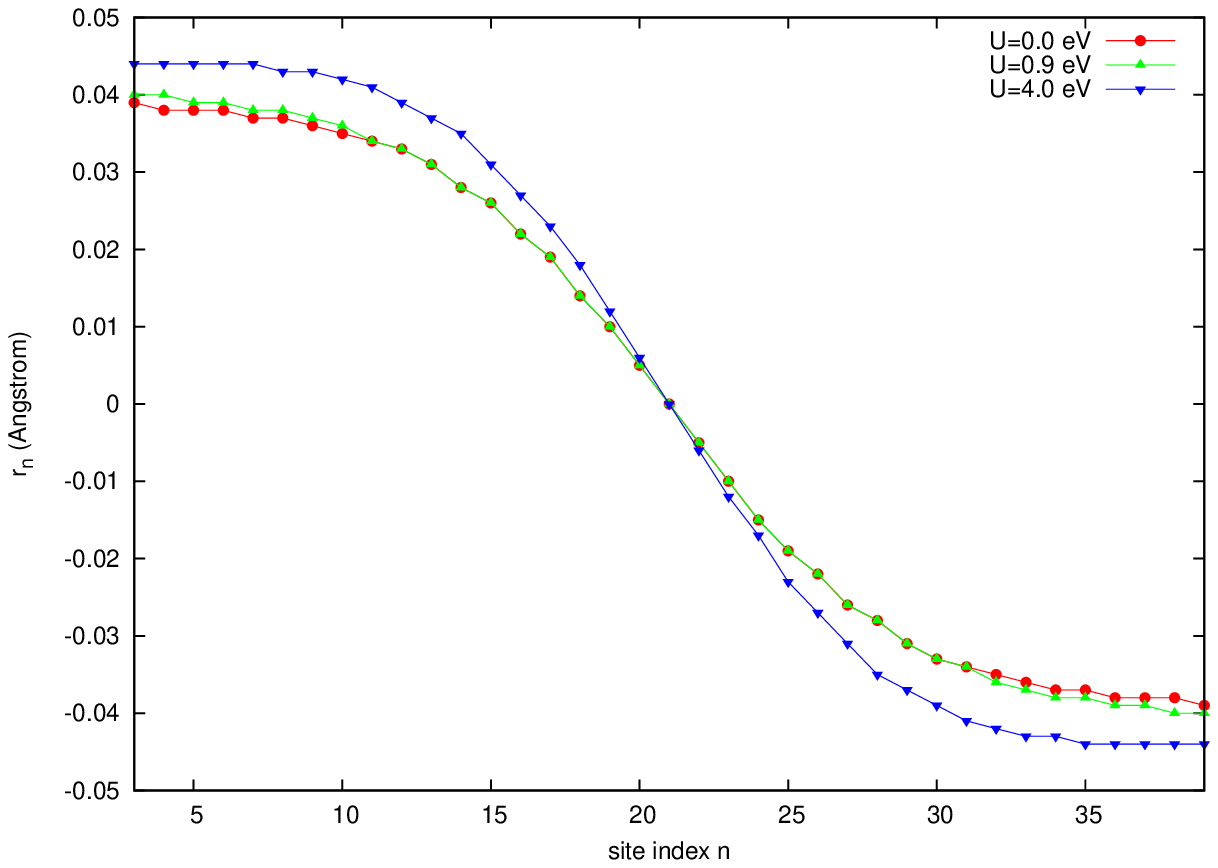}
\caption{\label{fig:g_U} The staggered bond order parameter $r_n$ of
a static charged soliton for several different $U$ vaslues. ($\epsilon_{\rho}=5.0\times10^{-7}$, V=0.0 eV)}
\end{figure}
Firstly, we study the Hubbard model with only on-site Coulomb interactions $U$, \textit{i.e.}, $V=0$ in
the EHM. The dependence of the soliton stationary velocity $v_s$ on
the on-site Coulomb interactions $U$ is displayed in
Fig.~\ref{fig:v_U}. Interestingly, $v_s$ is non-monotonic in $U$: it
grows to a shallow maximum at $U\approx0.9$ eV. Actually, the increase or decrease of $v_s$ is strongly related to the degree of delocalization of the soliton defect. In Fig.~\ref{fig:c_U} we show the net charge distribution pictures of a static charged soliton calculated with different $U$ vaslues. Clearly, there are large charge fluctuations within the defect central part due to the existence of a soliton. While the on-site repulsion $U$ increases, a large charge fluctuation becomes energetically unfavorable and the net charge has the tendency to be equally distributed at different CH monomers. Therefore, it can be clearly seen from Fig.~\ref{fig:c_U} that, the charge oscillation is reduced substantially with the increase of $U$. In fact, at the early stage of the increase of
the on-site Coulomb interaction $U$ (when the values of $U$ are not too large), this increase is in favor of the transport of the soliton because the on-site repulsion has the tendency to push the electron (or hole) in the accumulated large charge density of the central part of a charged soliton to hop more easily to the neighbor site. In other words, within this stage the delocalization of the soliton defect is strengthened and consequently soliton transport becomes easier. Therefore, the stationary velocity of a charged soliton $v_s$ increases while the value of $U$ increases from 0.0 eV to 0.9 eV as displayed in Fig.~\ref{fig:v_U}. On the other hand,
when the on-site Coulomb interactions $U$ increase further and
become to be dominant, the lattice tends to be occupied by one
electron per site. This means the electrons are more localized and
accordingly the free motion of the charged soliton will be greatly
restricted by large on-site repulsion $U$, so the delocalization of the soliton defect is decreased and consequently the transport of the soliton becomes more difficult. Therefore, $v_s$ decreases while the value of $U$ increases further from 0.9 eV to 4.0 eV as shown in Fig.~\ref{fig:v_U}. The change of delocalization level of the charged soliton defect can be directly viewed through the geometrical picture of the static soliton defect. In Fig.~\ref{fig:g_U}, the staggered bond order
parameter $r_n$ of a static charged soliton calculated by different $U$ values is
shown. The geometrical distortion of soliton defect from the regular
single and double bond alternation is normally described by a
hyperbolic tangent function,\cite{Su79, Su80}
\begin{equation}
\label{eq1} r_n =r_\infty \tanh \left( {\frac{n-C} {L}} \right).
\end{equation}
where, $r_\infty$ is the staggered bond order parameter in infinite
\textit{trans}-PA chain without defect, $C$ denoting the defect
center, and $L$ is the half-width of the soliton, determined in the
way that it will give the best fitting of $r_n$ against $n$. We extract the soliton width $L$ as:
7.95 ($U=0.0 $ eV), 8.18 ($U=0.9$ eV) and 6.64 ($U=4.0 $ eV). These
data certainly show that the delocalization of the soliton defect is
enhanced with the increase of $U$ from $0.0 $ eV to $0.9$ eV and then
reduced substantially with the increase of $U$ from $0.9$ eV to $4.0 $ eV.
This sequence is in good accordance with the sequence for the
stationary velocity of a charged soliton $v_s$ illustrated in Fig.~\ref{fig:v_U}.

\begin{figure}
\includegraphics[width =8 cm]{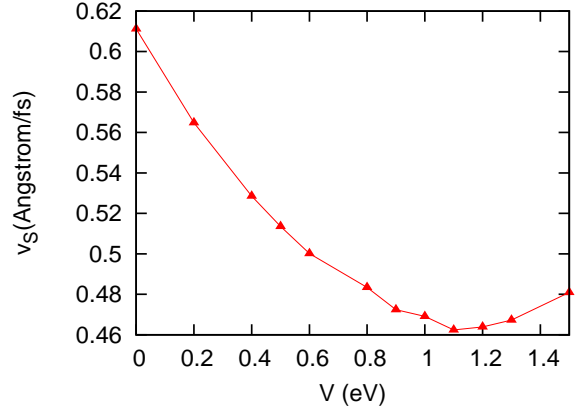}
\caption{\label{fig:v_V} The stationary velocity of a charged soliton as a function of different $V$ values. ($\Delta t$=0.05 fs, $\epsilon_{\rho}=5.0\times10^{-7}$, $E_0$=6.0 mV/\AA, U=0.0 eV)}
\end{figure}
\begin{figure}
\includegraphics[width =14 cm]{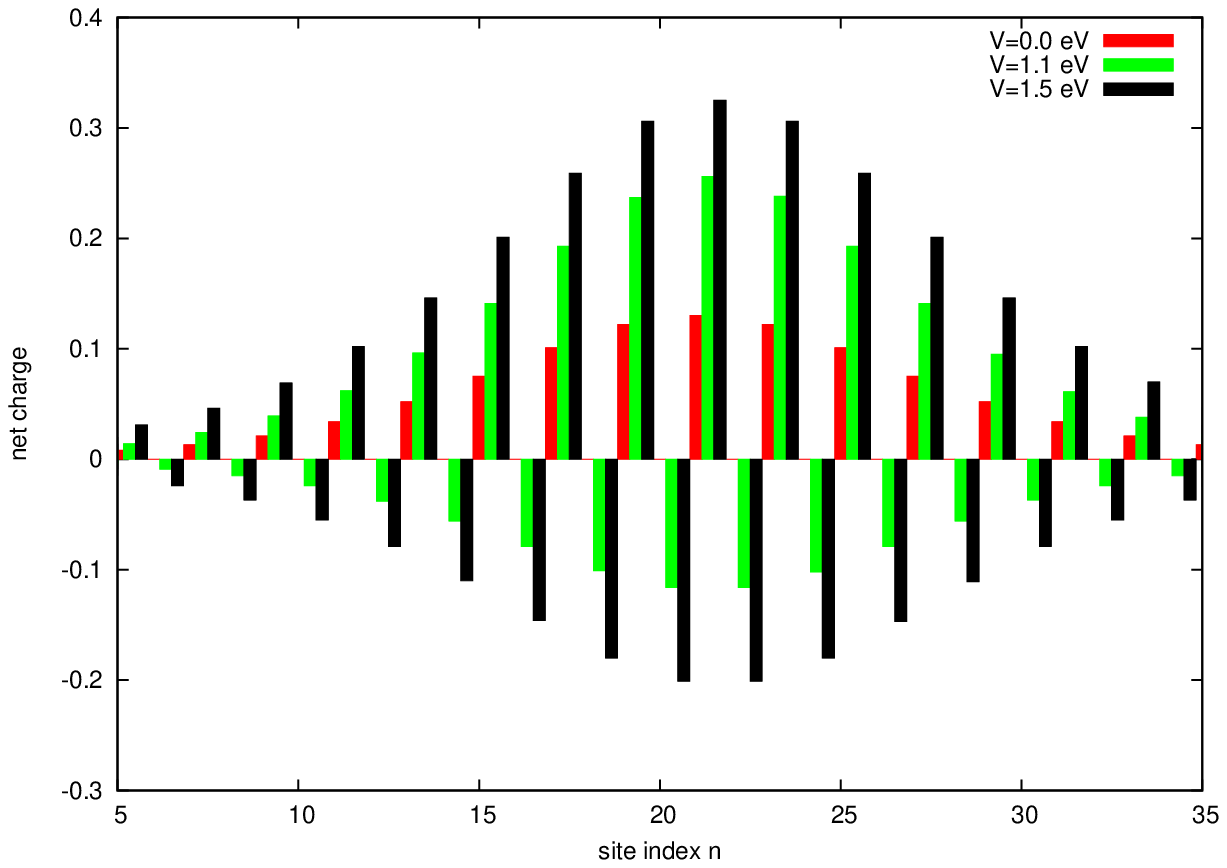}
\caption{\label{fig:c_V} The net charge of a static charged soliton as a function of the site
index calculated with different $V$ values. ($\epsilon_{\rho}=5.0\times10^{-7}$, U=0.0 eV)}
\end{figure}
\begin{figure}
\includegraphics[width =14 cm]{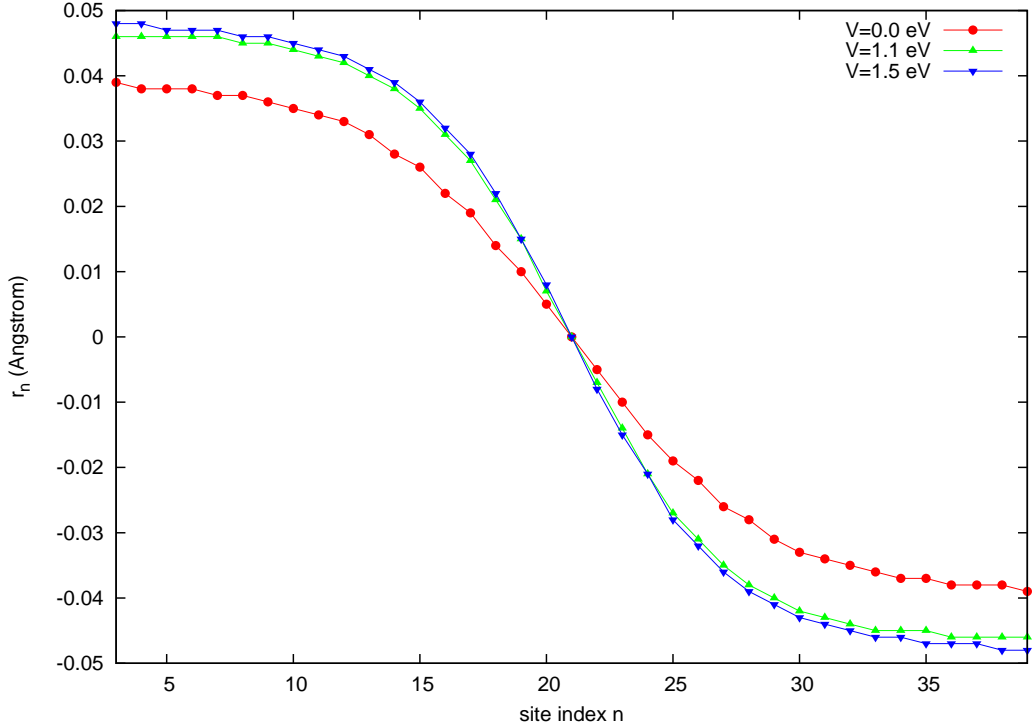}
\caption{\label{fig:g_V} The staggered bond order parameter $r_n$ of
a static charged soliton for several different $V$ values. ($\epsilon_{\rho}=5.0\times10^{-7}$, U=0.0 eV)}
\end{figure}
Secondly, we also study the influence of nearest neighbor
electron-electron interactions $V$ on the charged soliton transport process.
The dependence of the stationary velocity of a charged soliton $v_s$ on the $V$
values is displayed in Fig.~\ref{fig:v_V}, where $U$ values are
supposed to be frozen to be zero. This assumption is not realistic
because the on-site Coulomb interactions $U$ are normally much
stronger than the neighbor electron-electron interactions $V$; we make this assumption only for the purpose of
studying of the influence of $V$ on the soliton transport process
without the effect of $U$. As can be seen in Fig.~\ref{fig:v_V}, the situation is completely different from the case of an on-site Coulomb interaction: $v_s$ goes through a minimum for finite $V$, not a maximum. Actually, similar to the the case discussed above with $U$ varying, the increase or decrease of $v_s$ with $V$ is also strongly related to the change of the delocalization abilities of the charged soliton defect. In Fig.~\ref{fig:c_V} we show the net charge distribution pictures of a static charged soliton calculated with different $V$ values. It can be clearly seen that the nearest neighbor
electron-electron interactions $V$ will induce negative charge densities at the neighboring sites of the positive charge density and the charge polarization will become larger and larger with increasing $V$. For small $V$, larger positive charge densities will be induced in the central part of a charged soliton defect while the induced negative charge densities at nearest neighbor sites are very small. Therefore the charged soliton defect tends to be more localized and consequently the charged soliton transport becomes more difficult. So, the velocity of the charged soliton $v_s$ decreases while the value of $V$ increases from 0.0 eV to 1.1 eV as displayed in Fig.~\ref{fig:v_V}. However, when nearest neighbor electron-electron interactions $V$ increase further and
become dominant, large charge polarization will be induced. Therefore, the electron-hole attractions between opposite charge densities at nearest-neighbor sites will contribute much more significantly to the soliton system and favor the hoppings of the accumulated electrons (or holes) in the central part of a charged soliton defect to the neighbor sites. This change leads to a more delocalized soliton defect and accordingly the increase
of the velocity of the charged soliton $v_s$.
In order to directly view the change of delocalization level of the charged soliton defect through geometrical pictures,
we also show the staggered bond order parameter $r_n$ of a static charged
soliton calculated with different $V$ values in Fig.~\ref{fig:g_V},. We make hyperbolic tangent
function fittings for these curves and obtain the soliton width $L$
for different $V$ values: 7.95 ($V=0.0 $ eV), 6.08 ($V=1.1 $ eV) and
6.19 ($V=1.5 $ eV). The change of the soliton width $L$ shows
the change of the delocalization of the charged soliton. This
sequence is in good agreement with the sequence of the stationary
velocity of a charged soliton $v_s$ calculated for different $V$ values illustrated in Fig.~\ref{fig:v_V}.

\begin{figure}
\includegraphics[width =14 cm]{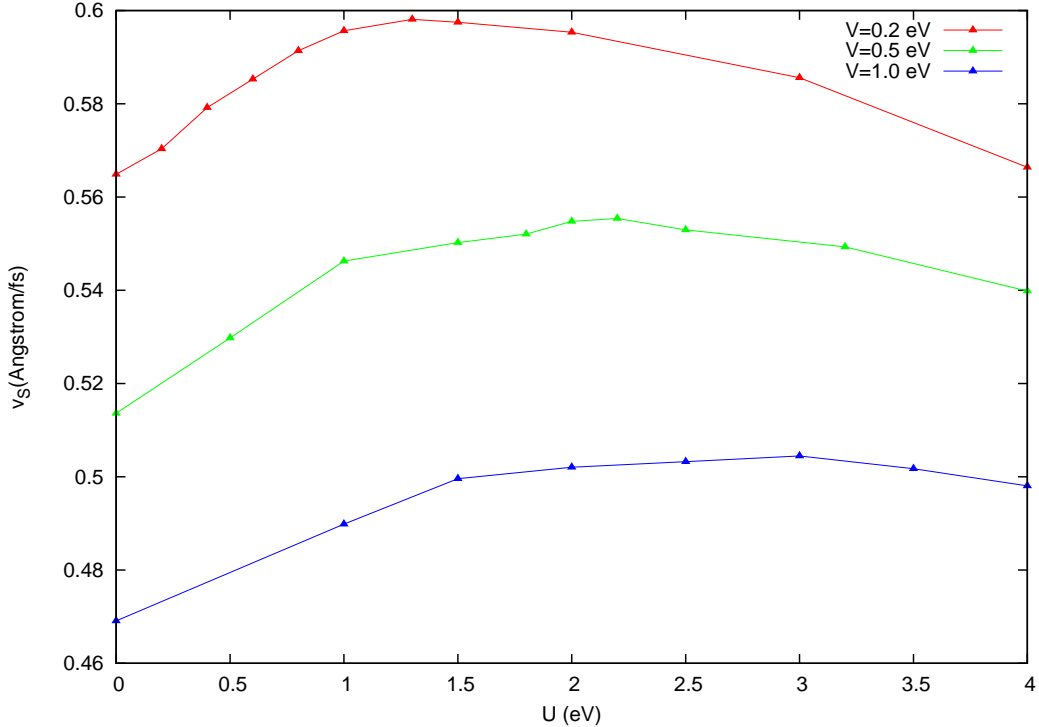}
\caption{\label{fig:v_f} The stationary velocity of a charged soliton as a function of $U$
for fixed values of $V$. ($\Delta t=0.05$ fs,
$\epsilon_{\rho}=5.0\times10^{-7}$, $E_0=3.0$ mV/\AA)}
\end{figure}
Furthermore, we consider the realistic case of \textit{trans}-PA in which both the on-site Coulomb
interactions $U$ and the nearest-neighbor interactions $V$ are taken into account.
Fig.~\ref{fig:v_f} shows the stationary velocity of a charged soliton $v_s$ as a
function of $U$ for fixed values of $V=0.2 $ eV, $V=0.5 $ eV and
$V=1.0 $ eV. Generally, the $v_s$ behavior with fixed $V$ values is quite similar to that has been illustrated in Fig.~\ref{fig:v_U} with fixed $V$=0.0 eV. It can be found that $v_s$ can be remarkably improved with increasing $U$
values at first and then gradually decreases with $U$ after a shallow maximum. This result can also be easily understood based on the simple analysis in HM. For a fixed value of $V$, as the on-site repulsions $U$ increase and gradually overcome the nearest-neighbor interactions $V$, the electron (or hole) will hop more easily to the next site, so the stationary velocity of a charged soliton increases. On the other hand, when the on-site repulsions $U$ increase further and become to be dominant, the lattice tends to be occupied by one electron per site. The soliton motion will be greatly prevented by $U$, so the stationary velocity of a charged soliton decreases. Additionally, the $U$ value where $v_s$ shows a shallow maximum increases while $V$ is increased.

\section{Summary and conclusion}
For a model conjugated polymer chain initially holding a charged
soliton defect, described using the combined SSH-EHM model extended
with an additional part for the influence of an external electric
field, we have solved the time-dependent Schr\"{o}dinger equation
for the $\pi$ electrons and the equations of motion for the monomer
displacements in a nonadiabatic way by virtue of
the combination of the adaptive TDDMRG and classical MD.

We find that after a smooth turn-on of the external electric field the charged soliton is accelerated at
first and then moves with a constant velocity as one entity consisting of both the charge and the lattice deformation. During the entire time
evolution process the geometrical distortion curve and charge
distribution shape for the charged soliton defect always stay coupled
and show no dispersion, implying that the soliton defect is an
inherent feature and the fundamental charge carrier in
\textit{trans}-PA. The dependence of the stationary velocity of a charged soliton $v_s$ on the external electric field
strength is also studied, and an ohmic region where $v_s$ increases linearly with
the field strength is found. This ohmic region extends approximately from 6 mV/$\text{\AA}$ to 12 mV/$\text{\AA}$ for the case of $U$=2.0 eV and $V$=1.0 eV. Below 6 mV/$\text{\AA}$  the soliton velocity increases nonlinearly and no saturation of $v_s$ with increasing electric field strength is found.

The influence of electron-electron interactions (both the on-site Coulomb
interactions $U$ and the nearest-neighbor interactions $V$) on charged soliton motion are
investigated in detail. In general, small $U$ values are beneficial to the solitonic motion because they have the tendency to push the accumulated electrons (or holes) in the defect central part to hop more easily to the neighbor site, and large $U$ values are unfavorable for the soliton motion because the lattice tends to be occupied by one electron per site due to the domination of on-site repulsions. Meanwhile, small $V$ values are not beneficial to the solitonic motion because they will induce a more localized defect distribution, and due to the induced large charge polarization accompanied with large nearest-neighbor attractions large $V$ values favor the soliton transport. When $U$ and $V$ are considered at the same time, $v_s$ can be significantly enhanced at the beginning stage of increasing $U$ and then decreases gradually with $U$ after a shallow maximum.

Because our calculations by the nonadiabatic
TDDMRG/MD method give interesting charged soliton transport
pictures and reveal the relationship between electron-electron
interactions and the transport of charged solitons while all
relevant electron-electron and electron-phonon interactions are
nearly fully taken into account, we consider that this combination
of methods provides a new nonadiabatic dynamic simulation method for exploring
the physical properties of electrons correlation systems with
relevant lattice dynamics.

\section*{Acknowledgment}
HM is grateful to Chungen Liu and Adrian Kleine for helpful
discussions. HM also acknowledges the support by Alexander von
Humboldt Research Fellowship.


\section*{References}
\begin{thebibliography}{}
\bibitem{Chiang77} 
C. K. Chiang, C. R. Fincher, Jr., Y. W. Park, A. J. Heeger, H. Shirakawa, E. J. Louis, S. C. Gau, and Alan G. MacDiarmid,
Phys. Rev. Lett. \textbf{39}, 1098 (1977).

\bibitem{Shirakawa77}
H. Shirakawa, E. J. Louis, Alan G. MacDiarmid, C. K. Chiang, and A. J. Heeger, J. Chem. Soc., Chem. Commun.
\textbf{16}, 579 (1977).

\bibitem{Chiang78} 
C. K. Chiang, M. A. Druy, S. C. Gau, A. J. Heeger, E. J. Louis, Alan G. MacDiarmid, Y. W. Park, and H. Shirakawa, J. Am. Chem.
Soc. \textbf{100}, 1013 (1978).

\bibitem{Heeger01} 
A. J. Heeger, J. Phys. Chem. B \textbf{105}, 18475 (2001).

\bibitem{Heeger01_2} 
A. J. Heeger, Rev. Mod. Phys. \textbf{73}, 681 (2001).

\bibitem{Baeriswyl92}
D. Baeriswyl, D. K. Campbell, and S. Mazumdar, \textit{Conjugated Conducting Polymers (H. Kiess. ed.)} (Springer-Verlag, Berlin, 1992).

\bibitem{Barford05}
W. Barford, \textit{Electronic and Optical Properties of Conjugated Polymers} (Oxford University Press, Oxford, 2005).

\bibitem{Su79} 
W. P. Su, J. R. Schrieffer, and A. J. Heeger, Phys. Rev.
Lett. \textbf{42}, 1698 (1979).

\bibitem{Su80}
W. P. Su, J. R. Schrieffer, and A. J. Heeger, Phys. Rev. B \textbf{22}, 2099 (1980).

\bibitem{Heeger88} 
A. J. Heeger, S. Kivelson, J. R. Schrieffer, and W. P. Su,
Rev. Mod. Phys. \textbf{60}, 781 (1988).

\bibitem{Bishop84}
A. R. Bishop, D. K. Campbell, P. S. Lomdahl, B. Horovitz, and
S. R. Phillpot, Synth. Met. \textbf{9}, 223 (1984).

\bibitem{Johansson02}
\AA. Johansson and S. Stafstr\"{o}m, Phys. Rev. B
\textbf{65}, 045207 (2002).

\bibitem{Forner88} 
W. F\"{o}rner, C. L. Wang, F. Martino, and J. Ladik, Phys.
Rev. B \textbf{37}, 4567 (1988).

\bibitem{Forner98} 
W. F\"{o}rner, and W. Utz, J. Mol. Model. \textbf{4}, 12 (1998).

\bibitem{Champagne97}
B. Champagne, E. Deumens, and Y. \"{O}hrn, J. Chem. Phys. \textbf{107}, 5433(1997).

\bibitem{Yonemitsu88} 
K. Yonemitsu, Y. Ono, and Y. Wada, J. Phys. Soc. Jpn. \textbf{57}, 3875 (1988).

\bibitem{Sim91} 
F. Sim, D. R. Salahub, S. Chin, and M. Dupuis, J. Chem.
Phys. \textbf{95}, 4317 (1991).

\bibitem{Suhai92} 
S. Suhai, Int. J. Quantum. Chem.
\textbf{42}, 193 (1992).

\bibitem{Villar92} 
H. O. Villar and M. Dupuis, Theor. Chim. Acta. \textbf{83}, 155 (1992).

\bibitem{Bally92} 
T. Bally, K. Roth, W. Tang, R. R. Schrock, K. Knoll, and L.
Y. Park, J. Am. Chem. Soc. \textbf{114}, 2440 (1992).

\bibitem{Rodriguez-Monge95} 
L. Rodr\'{i}guez-Monge and S. Larsson, J. Chem. Phys.
\textbf{102}, 7106 (1995).

\bibitem{Hirata95}
S. Hirata, H. Torii, and M. Tasumi, J. Chem. Phys. \textbf{103}, 8964 (1995).

\bibitem{Fulscher95} 
M. P. F\"{u}lscher, S. Matzinger, and T. Bally, Chem.
Phys. Lett. \textbf{236}, 167 (1995).

\bibitem{Guo97}
H. Guo and J. Paldus, Int. J. Quantum. Chem. \textbf{63}, 345 (1997).

\bibitem{Perpete99}
E. A. Perp\`{e}te and B. Champagne, J. Mol. Struct.
(THEOCHEM) \textbf{487}, 39 (1999).

\bibitem{Fonseca01} 
T. L. Fonseca, M. A. Castro, C. Cunha, and O. A. V. Amaral,
Synth. Met. \textbf{123}, 11 (2001).

\bibitem{Oliveira03} 
L. N. Oliveira, O. A. V. Amaral, M. A. Castro, and T. L. Fonseca,
Chem. Phys. \textbf{289}, 221 (2003).

\bibitem{Champagne04} 
B. Champagne and M. Spassova, Phys. Chem. Chem. Phys.
\textbf{6}, 3167 (2004).

\bibitem{Monev05}
V. Monev, M. Spassova, and B. Champagne, Int. J. Quantum.
Chem. \textbf{104}, 354 (2005).

\bibitem{White92}
S. R. White, Phys. Rev. Lett. \textbf{69},
2863 (1992).

\bibitem{White93}
S. R. White, Phys. Rev. B \textbf{48},
10345 (1993).

\bibitem{Schollwock05}
U. Schollw\"{o}ck, Rev. Mod. Phys. \textbf{77}, 259 (2005).

\bibitem{Pang95}
H. Pang, and S. Liang, Phys. Rev. B \textbf{51},
10287 (1995).

\bibitem{Wen96}
G. Z. Wen, and W. P. Su, Synth. Met. \textbf{78}, 195 (1996).

\bibitem{Lepetit97}
M. B. Leptit, and G. M. Pastor, Phys. Rev. B \textbf{56},
4447 (1997).

\bibitem{Yaron98}
D. Yaron, E. E. Moore, Z. Shuai, and J. L. Bredas, J. Chem. Phys.
\textbf{108}, 7451 (1998).

\bibitem{Boman98}
M. Boman, and R. J. Bursill, Phys. Rev. B \textbf{57}, 15167 (1998).

\bibitem{Kuwabara98}
M. Kuwabara, Y. Shimoi, and S. Abe, J. Phys. Soc. Jpn.
\textbf{67}, 1521 (1998).

\bibitem{Shuai98}
Z. Shuai, J. L. Bredas, A. Saxena, and A. R. Bishop, J. Chem. Phys.
\textbf{109}, 2549 (1998).

\bibitem{Shuai98_2}
Z. Shuai, J. L. Bredas, S. K. Pati, and S. Ramasesha, Phys. Rev. B
\textbf{58}, 15329 (1998).

\bibitem{Bursill99}
R. J. Bursill, and W. Barford, Phys. Rev. Lett. \textbf{82}, 1514 (1999).

\bibitem{Zhang00}
G. P. Zhang, Phys. Rev. B \textbf{61}, 4377 (2000).

\bibitem{Barford01}
W. Barford, and R. J. Bursill, Synth. Met. \textbf{119}, 251 (2001).

\bibitem{Barford01_2}
W. Barford, and R. J. Bursill, Phys. Rev. B \textbf{63}, 195108 (2001).

\bibitem{Race01}
A. Race, W. Barford, and R. J. Bursill, Phys. Rev. B \textbf{64}, 035208 (2001).

\bibitem{Barford02}
W. Barford, R. J. Bursill, and R. W. Smith, Phys. Rev. B \textbf{66}, 115205 (2002).

\bibitem{Barford02_2}
W. Barford, R. J. Bursill, and M. Y. Lavrentiev, Phys. Rev. B \textbf{65}, 075107 (2002).

\bibitem{Raghu02}
C. Raghu, Y. A. Pati, and S. Ramasesha, Phys. Rev. B \textbf{65}, 155204 (2002).

\bibitem{Race03}
A. Race, W. Barford, and R. J. Bursill, Phys. Rev. B \textbf{67}, 245202 (2003).

\bibitem{Ma04}
H. Ma, C. Liu, and Y. Jiang, J. Chem. Phys.
\textbf{120}, 9316 (2004).

\bibitem{Moore05}
E. E. Moore, W. Barford, and R. J. Bursill, Phys. Rev. B \textbf{71}, 115107 (2005).

\bibitem{Ma05}
H. Ma, F. Cai, C. Liu, and Y. Jiang, J. Chem. Phys.
\textbf{122}, 104909 (2005).

\bibitem{Ma05_2}
H. Ma, C. Liu, and Y. Jiang, J. Chem. Phys.
\textbf{123}, 084303 (2005).

\bibitem{Yan05}
Y. Yan, and S. Mazumdar, Phys. Rev. B \textbf{72}, 212201 (2005).

\bibitem{Ma06}
H. Ma, C. Liu, and Y. Jiang, J. Phys. Chem. B
\textbf{110}, 26488 (2006).

\bibitem{Hu07}
W. Hu, H. Ma, C. Liu, and Y. Jiang, J. Chem. Phys.
\textbf{126}, 044903 (2007).

\bibitem{Ma07}
H. Ma, C. Liu, and Y. Jiang, J. Phys. Chem. A
\textbf{111}, 9471 (2007).

\bibitem{White04}
S. R. White and A. Feiguin, Phys. Rev. Lett. \textbf{93}, 076401 (2004).

\bibitem{Daley04}
A. J. Daley, C. Kollath, U. Schollw\"{o}ck, and G. Vidal, J. Stat. Mech. P04005 (2004).

\bibitem{Ehrenfreund87}
E. Ehrenfreund, Z. Vardeny, O. Brafman, and B. Horovitz, Phys. Rev. B
\textbf{36}, 1535 (1987).

\bibitem{Rakhmanova99}
S. V. Rakhmanova, and E. M. Conwell, Appl. Phys. Lett. \textbf{75}, 1518 (1999).

\bibitem{Rakhmanova00}
S. V. Rakhmanova, and E. M. Conwell, Synth. Met. \textbf{110}, 37 (2000).

\bibitem{Johansson01}
\AA. Johansson and S. Stafstr\"{o}m, Phys. Rev. Lett.
\textbf{86}, 3602 (2001).

\bibitem{Johansson02}
\AA. Johansson and S. Stafstr\"{o}m, Phys. Rev. B
\textbf{65}, 045207 (2002).

\bibitem{Basko02} 
D. M. Basko, and E. M. Conwell, Phys. Rev. Lett. \textbf{88}, 056401 (2002).

\bibitem{Johansson04}
\AA. Johansson and S. Stafstr\"{o}m, Phys. Rev. B
\textbf{69}, 235205 (2004).

\bibitem{Fu06}
J. Fu, J. Ren, X. Liu, D. Liu, and S. Xie, Phys. Rev. B
\textbf{73}, 195401 (2006).

\bibitem{Zhao08}
H. Zhao, Y. Yao, Z. An, and C. Q. Wu, Phys. Rev. B \textbf{78}, 035209 (2008).

\bibitem{Ma08}
H. Ma and U. Schollw\"{o}ck, J. Phys. Chem. in revision. (arXiv:0810.2676)

\bibitem{Schollwock07}
U.  Schollw\"{o}ck, J. Magn. Mag. Mat. \textbf{310}, 1394 (2007).

\bibitem{Schollwock07_2}
U.  Schollw\"{o}ck, Int. J. Mod. Phys. B \textbf{21}, 2564 (2007).

\bibitem{Chadi78}
D. J. Chadi, Phys. Rev. Lett. \textbf{41},
1062 (1978).

\bibitem{Payne92}
M. C. Payne, M. P. Teter, D. C. Allan, T. A. Arias, J. D. Joannopoulos, Rev. Mod. Phys. \textbf{64}, 1045 (1992) and the references therein.

\bibitem{Hellmann37}
H. Hellmann, \textit{Einf\"{u}hrung in die Quantenchemie} (Franz Deuticke, Leipzig, 1937).

\bibitem{Feynman39}
R. P. Feynman, Phys. Rev. \textbf{56}, 340 (1939).

\bibitem{Vidal04}
G. Vidal, Phys. Rev. Lett. \textbf{93},
040502 (2004).

\bibitem{Suzuki76}
M. Suzuki, Prog. Theor. Phys. \textbf{56},
1454 (1976).

\bibitem{Klumper91}
A. Kl\"{u}mper, A. Schadschneider, and J. Zittartz, J. Phys. A: Math. Gen.
\textbf{24}, L955 (1991).

\bibitem{Fannes92}
M. Fannes, B. Nachtergaele, and R. F. Werner, Commun. Math. Phys. \textbf{144}, 443 (1992).

\bibitem{Klumper92}
A. Kl\"{u}mper, A. Schadschneider, and J. Zittartz, Z. Phys. B \textbf{87}, 281 (1992).

\bibitem{Derrida93}
B. Derrida, M. R. Evans, V. Hakim, and V. Pasquier, J. Phys. A: Math. Gen. \textbf{26}, 1493 (1993).

\bibitem{Ian07}
I. P. McCulloch, J. Stat. Mech. P10014 (2007).

\bibitem{Gobert05}
D. Gobert, C. Kollath, U. Schollw\"{o}ck, and G. Sch\"{u}tz, Phys. Rev. E \textbf{71}, 036102 (2005).

\bibitem{Kollath05}
C. Kollath, U. Schollw\"{o}ck, and W. Zwerger, Phys. Rev. Lett. \textbf{95}, 176401 (2005).

\bibitem{Kleine08}
A. Kleine, C. Kollath, I. P. McCulloch, T. Giamarchi, and U. Schollw\"{o}ck, Phys. Rev. A \textbf{77}, 013607 (2008).

\bibitem{Cramer08}
M. Cramer, A. Flesch, I. P. McCulloch, U. Schollw\"{o}ck, and J. Eisert, Phys. Rev. Lett. \textbf{101}, 063001 (2008).

\bibitem{Feiguin05}
A. E. Feiguin and S. R. White, Phys. Rev. B \textbf{72}, 020404 (2005).



\end{thebibliography}
\end{document}